\documentclass[conference]{IEEEtran}
\IEEEoverridecommandlockouts

\usepackage{cite}
\usepackage{amsmath,amssymb,amsfonts}
\usepackage{algorithmic}
\usepackage{graphicx}
\usepackage{xcolor}
\usepackage{booktabs}
\usepackage{multirow}
\usepackage{array}
\usepackage{enumitem}
\usepackage{stfloats}

\def\BibTeX{{\rm B\kern-.05em{\sc i\kern-.025em b}\kern-.08em
    T\kern-.1667em\lower.7ex\hbox{E}\kern-.125emX}}

\begin{document}

\title{Energy Trading Potential Index for a Peer-to-Peer\\
Smart Grid Community with Flexible Prosumer Role Switching}

\author{
\IEEEauthorblockN{Zain Imran, Sana Humayun, Muhammad Shahzaib Saleem, Naveed Ul Hassan}
\IEEEauthorblockA{Lahore University of Management Sciences (LUMS), Lahore, Pakistan 54792\\
Emails: zain.imran@lums.edu.pk, 26100273@lums.edu.pk, muhammad.shahzaib@lums.edu.pk, naveed.hassan@lums.edu.pk}
}

\maketitle

\begin{abstract}
In many electricity markets, declining feed-in tariffs have made grid export increasingly unattractive for residential solar prosumers, while retail 
electricity prices remain high.
Peer-to-peer (P2P) energy trading offers a direct alternative, but it requires a dedicated infrastructure layer for real-time bilateral matching, automated settlement, and tamper-proof transaction records, for which blockchain is widely 
proposed. Deploying such infrastructure must be economically justified by the 
community's actual trading potential. A critical and underexplored question is 
whether trading potential survives as communities become prosumer-heavy, since under fixed role assignment all households eventually end up on the supply side with no buyers remaining. This paper addresses these gaps by proposing the Energy Trading Potential Index (ETPI), a normalized data-driven metric that quantifies the structural impact of flexible role switching on community-level trading potential, where prosumers dynamically join the buyer side whenever they are in energy deficit. The P2P market is modeled as a generalized bipartite graph and pairwise interaction scores aggregated over trading rounds compute the ETPI in [0,\,1]. Simulation results using the PRECON residential dataset and NREL PVWatts solar profiles show that for the (1:9) prosumer-heavy mix, the flexible policy achieves an ETPI of 0.61 versus only 0.15 under the static policy, a fourfold improvement that the static model entirely misses. The ETPI framework serves as a lifecycle decision-support tool for evaluating and monitoring P2P energy trading infrastructure.
\end{abstract}


\section{Introduction}
\label{sec:intro}

Energy systems worldwide are undergoing a fundamental structural shift. The rapid 
decline in the cost of rooftop solar photovoltaic (PV) panels has made distributed 
energy generation economically accessible to ordinary households, transforming passive 
consumers into \emph{prosumers}, which are entities that both produce and consume 
electrical energy \cite{soto2021, tushar2023}. In many electricity markets, feed-in 
tariffs for residential solar export have declined significantly, with global data 
showing that feed-in tariffs and premiums have dropped from being the dominant 
procurement mechanism to representing just 10\% of renewable energy growth 
\cite{iea2025}. This growing spread between buyback rates and retail electricity 
prices makes selling surplus energy back to the grid increasingly unattractive for 
prosumers. Peer-to-peer (P2P) energy trading offers a direct alternative, enabling 
surplus-holding prosumers to trade directly with energy-deficient neighbors at 
mutually beneficial prices. Unlike net metering, P2P trading requires a dedicated 
infrastructure layer capable of real-time bilateral matching, automated settlement 
without a utility intermediary, and tamper-proof transaction records \cite{tanis2025}.

Blockchain technology is widely proposed to deliver these infrastructure properties
through distributed consensus, cryptographic data integrity, and automated
smart-contract settlement \cite{bhavana2024, wang2019, dorri2019, brooklyn2018,
nazari2024, khorasany2021, alskaif2022}.
Despite these advantages, blockchain carries substantial computational and operational
overhead, and that overhead can exceed the benefits in communities where trading
activity is low \cite{zade2022}.
Advanced P2P infrastructure should therefore only be deployed where trading potential
is sufficiently high to justify the cost --- yet the literature offers no principled,
data-driven metric for making that determination.
Works demonstrating blockchain-based P2P systems \cite{dorri2019, brooklyn2018,
nazari2024, khorasany2021} uniformly assume the community is already suitable for
trading, leaving the prior question of \emph{whether} it is suitable entirely
unanswered.
Assessments of trading potential in the residential sector have largely been confined
to technology and tariff availability scenarios \cite{neves2020}, without accounting
for the structural composition of the community itself.
Meanwhile, the broader P2P trading literature has examined market designs
\cite{tushar2020overview, tushar2023, tanis2025}, game-theoretic pricing among
prosumers \cite{paudel2019, lee2015}, and grid-level constraints
\cite{ullah2021tsg}, yet none of these works treats community readiness as a
prerequisite for infrastructure deployment.
A closely related gap concerns prosumer role flexibility.
In practice, a prosumer's role as buyer or seller changes dynamically each round
according to its instantaneous net energy balance \cite{tushar2020overview,
paudel2019, liu2020tii}, yet no prior work has quantified how much trading potential
is unlocked or suppressed by the role-assignment policy.
Bipartite matching frameworks \cite{zeng2021} similarly treat role assignment as a given rather than a design
variable, overlooking how policy choices shape the aggregate supply-demand structure
of the community.

This paper addresses these gaps by proposing the Energy Trading Potential Index 
(ETPI), a normalized data-driven metric that quantifies the structural impact of 
flexible role switching on community-level P2P trading potential. The P2P market is 
modeled as a generalized bipartite graph, with edge weights derived from pairwise 
interaction scores combining price compatibility, demand-supply balance, and 
communication link reliability. The aggregate of all edge weights gives the Energy 
Trading Potential (ETP), normalized into the ETPI in $[0, 1]$. Central to the 
framework is a \emph{flexible role-switching policy}, under which a prosumer's market 
role is determined dynamically each round by its instantaneous net energy balance, 
in contrast to a \emph{static policy} that permanently fixes prosumers on the supplier 
side. The ETPI can be seeded from historical consumption and generation data before 
deployment and continuously updated as real trading interactions accumulate, evolving 
from a pre-deployment estimate into a live monitoring instrument that tracks whether 
trading potential justifies sustained infrastructure costs over the system lifecycle.

The framework is validated through one-year simulations using real residential 
consumption data from the PRECON dataset \cite{nadeem2019} and NREL PVWatts solar 
profiles across ten consumer-to-prosumer mix configurations. Under the static policy, 
the balanced mix of five consumers and five prosumers (5:5) yields the highest ETPI 
of 0.42, while prosumer-heavy mixes score poorly due to the absence of buyer-side 
prosumers. Under the flexible policy, the mix of one consumer and nine prosumers (1:9) 
yields the best ETPI of 0.61, compared to only 0.15 for the same mix under the static 
policy. This fourfold improvement shows that communities with high solar penetration 
retain substantial trading potential when role flexibility is accounted for, directly 
answering whether trading activity collapses as neighborhoods become prosumer-heavy. 
The ETPI thus provides a principled pre-deployment signal for whether a community's 
trading activity justifies advanced P2P infrastructure investment. Sensitivity 
analysis further reveals that ETP is non-monotonically related to PV capacity in 
prosumer-heavy communities, meaning that maximizing installed solar capacity does not 
necessarily maximize trading potential.

The remainder of the paper is organized as follows. 
Section~\ref{sec:system} describes the system model.
Section~\ref{sec:interactions} describes the interaction model.
Section~\ref{sec:flexible} develops the flexible role-switching ETPI.
Section~\ref{sec:static} formulates the static policy as a special case and compares 
it to the flexible model.
Section~\ref{sec:simulations} presents and discusses the simulation methodology and 
results.
Section~\ref{sec:conclusion} concludes the paper.

\section{System Model}
\label{sec:system}

\subsection{Community Structure}

We consider a P2P energy trading community of $N$ residential households divided 
into two groups. Let $N_C$ denote the number of \emph{fixed consumers}, which are 
households with no local generation capability that always participate in the market 
as buyers. Let $N_P$ denote the number of \emph{prosumers}, which are households 
equipped with rooftop solar PV systems that can potentially supply surplus energy to 
others. We have $N_C + N_P = N$. No energy storage is assumed in the system. Any 
consumer demand not fulfilled through P2P trading is met by drawing from the utility 
grid at the prevailing retail tariff.

We define the \emph{community mix} as the tuple $(N_C : N_P)$, which characterizes 
the composition of the community. A perfectly balanced mix has $N_C = N_P$, while 
skewed mixes reflect a dominance of one group over the other. The community mix 
governs the structure of the trading graph and has a significant impact on the ETPI, 
as demonstrated through analysis and simulation in the following sections.

\subsection{Trading Rounds}

The trading horizon is divided into discrete \emph{trading rounds} indexed by $r$, 
each spanning one hour. The market operates only during daylight hours when solar 
generation is available. Each prosumer $P_j$ is equipped with a solar PV system of 
fixed rated capacity. In round $r$, its gross solar generation is $g_j^r$ (kWh) and 
household demand is $d_j^r$ (kWh), yielding a \emph{net energy balance}:
\begin{equation}
    s_j^r = g_j^r - d_j^r
    \label{eq:balance}
\end{equation}
If $s_j^r > 0$, the prosumer holds a surplus and participates as a supplier. If 
$s_j^r \leq 0$, the prosumer is in deficit and joins the demand side. This 
distinction underlies the flexible and static policies described in 
Sections~\ref{sec:flexible} and~\ref{sec:static}.

\begin{figure*}[!t]
    \centering
    \includegraphics[width=1\linewidth]{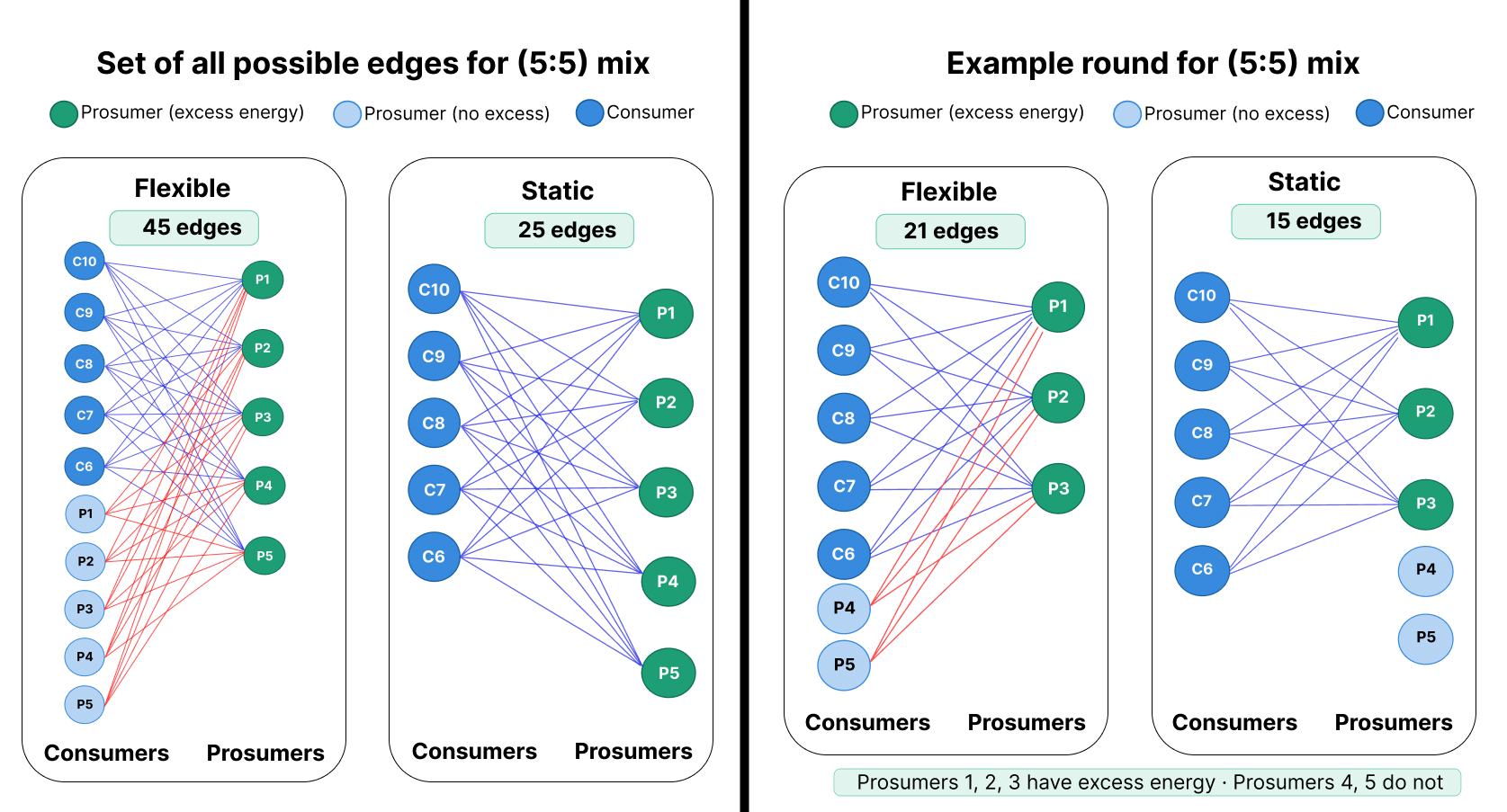}
    \caption{Set of all possible pairs for (5:5) mix under both flexible and static policy. Only 3 of the 5 prosumers have surplus energy in the example round. Blue edges represent consumer-prosumer pairs ($T_{ij}^r$), whereas red edges represent prosumer-prosumer pairs ($T_{kj}^r$)}. 
    \label{fig:network}
\end{figure*}

In each round $r$, consumer $C_i$ submits a \emph{bid price} $b_i^r$ (PKR/kWh) and 
\emph{demand} $\tilde{d}_i^r$ (kWh). Prosumer $P_j$, when $s_j^r > 0$, submits an 
\emph{offer price} $o_j^r$ (PKR/kWh) and \emph{surplus energy} $s_j^r$ (kWh). 
On the other hand, a prosumer $P_k$, when $s_k^r < 0$ submits a \emph{bid price} $b_k^r$ (PKR/kWh) and \emph{demand} $\tilde{d}_k^r$ (kWh) like a consumer. 
A complete bipartite graph is formed each round as shown in Figure~\ref{fig:network}, 
with consumers (including deficit prosumers) on the left and surplus prosumers on the 
right. Each edge carries a label of $+1$, $0$, or $-1$ reflecting the outcome of the 
pairwise interaction. A label of $+1$ denotes strong trading potential from favorable 
price alignment, sufficient surplus, and reliable links. A label of $-1$ denotes a 
non-viable trade due to price mismatch, insufficient surplus, or poor link 
reliability. A label of $0$ marks a neutral pair at the margin of viability.

\section{Interaction Model and Edge Weight Computation}
\label{sec:interactions}

The core of the ETPI framework is a pairwise interaction model that evaluates, 
for every potential consumer-supplier pair in every trading round, the feasibility 
and quality of a trade. The edge labels $+1$, $0$, and $-1$ introduced in 
Section~\ref{sec:system} correspond to the Positive, Neutral, and Negative 
sentiment categories defined formally in this section. We identify three key trading 
factors that govern this quality, combine them into an interaction function, and 
accumulate classification outcomes over time to compute edge weights.

\subsection{Trading Factors}

\textbf{Price Compatibility} ($W_{ij}^r$): A trade between consumer $C_i$ and 
supplier $P_j$ in round $r$ is price-feasible only if the consumer is willing to pay 
at least as much as the supplier demands. We define the price compatibility factor as
\begin{equation}
    W_{ij}^r = b_i^r - o_j^r
    \label{eq:price_compat}
\end{equation}
A trade is price-feasible only if $W_{ij}^r \geq 0$. The magnitude of $W_{ij}^r$ 
captures the degree of price alignment or mismatch between the pair. Whether positive 
or negative, the value is retained in full as an input to the interaction function 
rather than merely its sign.

\textbf{Demand Fulfillment} ($F_{ij}^r$): This factor reflects the energy 
compatibility between the supplier and the consumer, defined as the difference between 
the supplier's available surplus and the consumer's demand.
\begin{equation}
    F_{ij}^r = s_j^r - \tilde{d}_i^r
    \label{eq:demand_fulfill}
\end{equation}
A positive value ($F_{ij}^r \geq 0$) indicates that the supplier can fully satisfy 
the consumer's energy requirement. A negative value indicates a shortfall where the 
supplier's surplus is insufficient to cover the consumer's full demand.

\textbf{Network Robustness} ($R_{ij}^r$): Each consumer-prosumer pair $(i,j)$ is 
connected via a communication link with reliability $R_{ij}^r \in [0,1]$, drawn 
independently for each round. A value of $R_{ij}^r = 0$ indicates total link failure 
while $R_{ij}^r = 1$ indicates perfect connectivity. Poor link quality reduces the 
effective trading potential even when price and energy conditions are otherwise 
favorable. This factor serves as a conditional weight whose role in the interaction 
function changes depending on whether the energy condition is satisfied.

\subsection{Interaction Function}

The interaction function $f(\cdot)$ combines the three factors into a single 
real-valued score quantifying the trading potential for pair $(i,j)$ in round $r$. 
Network robustness acts as a conditional weighting factor. When the supplier can 
fulfill the consumer's demand ($F_{ij}^r \geq 0$), higher robustness amplifies the 
positive contribution of surplus. When the supplier cannot fully meet demand 
($F_{ij}^r < 0$), the reciprocal of robustness acts as a penalty where poor link 
quality compounds the energy shortfall. A small constant $\epsilon > 0$ (e.g., $\epsilon = 0.001$) is introduced to prevent 
division by zero, representing a minimum residual link quality below which the link 
is considered effectively failed. The interaction function is:
\begin{equation}
    f\!\left(W_{ij}^r,\, F_{ij}^r,\, R_{ij}^r\right) =
    \begin{cases}
        W_{ij}^r + R_{ij}^r \cdot F_{ij}^r, & \text{if } F_{ij}^r \geq 0 \\[6pt]
        W_{ij}^r + \dfrac{1}{R_{ij}^r + \epsilon} \cdot F_{ij}^r, & \text{if } F_{ij}^r < 0
    \end{cases}
    \label{eq:interaction}
\end{equation}

A positive score indicates a feasible and favorable interaction, while a negative 
score indicates an unfavorable or infeasible one. The output of \eqref{eq:interaction} 
is a real-valued continuous score with no fixed range. To normalize it, we apply the 
hyperbolic tangent:
\begin{equation}
    \bar{f}_{ij}^r = \tanh\!\left(f\!\left(W_{ij}^r,\, F_{ij}^r,\, R_{ij}^r\right)\right)
    \label{eq:tanh}
\end{equation}

The function $\tanh(\cdot)$ maps $\mathbb{R}$ into $(-1, 1)$, where values close to 
$+1$ indicate highly favorable interactions, values close to $-1$ indicate poor or 
failed interactions, and values near $0$ indicate neutral or marginal conditions. When 
interaction magnitudes are very large due to high price scales, most values saturate 
toward $\pm 1$, reducing sensitivity. In such cases, a robust normalization based on 
the median and interquartile range (IQR) can be used instead, ensuring that the full 
dynamic range of $(-1,1)$ is utilized regardless of the absolute scale of the 
interaction values.
\begin{equation}
    \tilde{f}_{ij}^r = \tanh\!\left(\frac{f_{ij}^r - \text{median}}{0.5 \cdot 
    \text{IQR}}\right)
    \label{eq:tanh_robust}
\end{equation}
For notational brevity, the indices $(i,j)$ are used as a generic pair reference throughout this section. For prosumer-prosumer pairs, the same 
equations apply with indices $(k,j)$ where $k$ denotes the deficit prosumer 
acting as consumer and $j$ denotes the surplus prosumer acting as supplier.

To categorize each normalized score, we define two thresholds $\theta_1 < 0 < \theta_2$ and assign one of three sentiment labels to each interaction:
\begin{equation}
    \text{Sentiment}_{ij}^r =
    \begin{cases}
        \text{Negative}, & \text{if } \bar{f}_{ij}^r \leq \theta_1 \\[2pt]
        \text{Neutral},  & \text{if } \theta_1 < \bar{f}_{ij}^r < \theta_2 \\[2pt]
        \text{Positive}, & \text{if } \bar{f}_{ij}^r \geq \theta_2
    \end{cases}
    \label{eq:classification}
\end{equation}

This three-way classification allows interaction histories to be accumulated compactly over long time windows, retaining only counts rather than full continuous scores for every pair and round.

There are two types of trading pairs. Consumer-prosumer pairs $(C_i, P_j)$ are active in any round where prosumer $P_j$ holds a surplus. Prosumer-prosumer pairs $(P_k, P_j)$ are active in rounds where $P_j$ holds a surplus and $P_k$ is in deficit, having switched to the consumer role under the flexible policy.

ETP is computed over a sliding window of $W = 11$ rounds, corresponding to one day. Each window produces one ETP value, and these are averaged across all windows to yield the final ETP for a given mix. The first step is computing the edge weight for each trading pair over the window. Figure ~\ref{fig:network} demonstrates all possible pairs for (5:5) mix for both flexible and static approach. The quantities $H_{ij}^r$ and $H_{kj}^r$ denote the number of rounds within the window in which the respective pair was active. For instance, $H_{ij}^r$ counts how many of the 11 rounds saw house $i$ acting as a consumer and house $j$ as a surplus-holding prosumer. As more windows are observed, the averaged ETP converges toward a more reliable characterization of the community's trading behavior.

For consumer-prosumer pairs, let $\alpha_{ij}^{CP}$, $\beta_{ij}^{CP}$, and $\delta_{ij}^{CP}$ denote the number of active rounds within the window that were classified as positive, neutral, and negative respectively, where $i$ is the fixed consumer and $j$ is the surplus-holding prosumer. For prosumer-prosumer pairs, the analogous counts $\alpha_{kj}^{PP}$, $\beta_{kj}^{PP}$, and $\delta_{kj}^{PP}$ apply, with $k$ as the deficit prosumer acting as consumer and $j$ as the surplus prosumer. Rounds in which a pair is inactive do not increment any counter. Negative interactions implicitly bring down the edge weight by increasing the denominator $H_{ij}^r$ while keeping the numerator same.

The edge weights for the two pair types are then defined as:
\begin{equation}
    T_{ij}^{CP} = \frac{\alpha_{ij}^{CP} + \gamma\,\beta_{ij}^{CP}}{H_{ij}^r}, 
    \qquad T_{ij}^{CP} \in [0, 1]
    \label{eq:edge_weight_cp}
\end{equation}
\begin{equation}
    T_{kj}^{PP} = \frac{\alpha_{kj}^{PP} + \gamma\,\beta_{kj}^{PP}}{H_{kj}^r}, 
    \qquad T_{kj}^{PP} \in [0, 1]
    \label{eq:edge_weight_pp}
\end{equation}
where $\gamma \in [0, 1]$ is a hyperparameter controlling the contribution of neutral interactions to the edge weight. Setting $\gamma = 0$ treats neutral interactions as non-contributory, while $\gamma = 1$ treats them as fully equivalent to positive ones. Intermediate values allow context-aware tuning depending on whether neutral outcomes reflect marginal but genuine trades or genuine infeasibility.
\section{Flexible Role-Switching}
\label{sec:flexible}

Flexible role-switching determines a prosumer's market role dynamically each 
trading round based on its instantaneous net energy balance \eqref{eq:balance}. 
If $s_j^r > 0$, prosumer $P_j$ acts as a supplier in round $r$. If 
$s_k^r \leq 0$, prosumer $P_k$ switches to the consumer role, submits a bid 
price $b_k^r$ drawn from the consumer price range, and its unmet demand is 
$|s_k^r| = d_k^r - g_k^r$. Letting $m^r$ denote the number of 
surplus-holding prosumers in round $r$, the effective community composition 
becomes:
\begin{align}
    \text{Consumers in round } r &= N_C + (N_P - m^r) = N - m^r \label{eq:flex_cons} \\
    \text{Suppliers in round } r &= m^r \label{eq:flex_pros}
\end{align}

\subsection{Active Edges}

Since trades can only occur between a supplier and a consumer, the number of active 
trading edges in round $r$ is:
\begin{equation}
    E_{\text{flex}}^r = m^r \cdot (N - m^r)
    \label{eq:eflex}
\end{equation}
This is a concave quadratic function in $m^r$, with boundary cases confirming expected 
behavior: $E_{\text{flex}}^r = 0$ when $m^r = 0$ (no suppliers) or $m^r = N$ 
(no consumers). When prosumer $P_j$ switches to the consumer role, it becomes 
eligible to trade with any surplus-holding prosumer $P_k$, creating prosumer-prosumer 
edges entirely absent in the static model. The total number of possible edges across all rounds is:
\begin{equation}
    E_{\text{flex}}^{\text{total}} = N_P(N - 1)
    \label{eq:etotal}
\end{equation}
which decomposes as:
\begin{equation}
    N_P(N-1) = \underbrace{N_C\,N_P}_{\text{consumer-prosumer pairs}} +
    \underbrace{N_P(N_P - 1)}_{\text{prosumer-prosumer pairs}}
    \label{eq:etotal_decomp}
\end{equation}
The second term quantifies the additional trading relationships that flexible 
switching unlocks. For a prosumer-heavy community $(1{:}9)$ with $N=10$, this yields 
$9 \times 9 = 81$ possible edges under the flexible policy versus only $1 \times 9 = 9$ 
under the static policy, a ninefold expansion.

\subsection{ETP under Flexible Policy}

For the trading pair consisting of deficit prosumer $P_k$ acting as consumer and 
surplus prosumer $P_j$ acting as supplier, the three trading factors are:
\begin{align}
    W_{kj}^r &= b_k^r - o_j^r \label{eq:w_flex}\\[2pt]
    F_{kj}^r &= s_j^r - |s_k^r| = s_j^r - (d_k^r - g_k^r) \label{eq:f_flex}\\[2pt]
    R_{kj}^r &\in [0,1] \label{eq:r_flex}
\end{align}
The interaction function \eqref{eq:interaction}, tanh normalization \eqref{eq:tanh}, 
and sentiment classification \eqref{eq:classification} apply to these values without 
modification, and the edge weight $T_{kj}^{PP}$ is computed from 
\eqref{eq:edge_weight_pp} over all rounds. The ETP under the 
flexible policy is:
\begin{equation}
    \text{ETP}_{\text{flex}} = 
    \sum_{i=1}^{N_C}\sum_{j=1}^{N_P} T_{ij}^{CP} +
    \sum_{k=1}^{N_P}\sum_{\substack{j=1 \\ j \neq k}}^{N_P} T_{kj}^{PP}
    \label{eq:etp_flex}
\end{equation}
where the first term aggregates all consumer-prosumer pair weights and the second 
term aggregates all prosumer-prosumer pair weights enabled by role switching. This ETP is calculated over a fixed window of W=11 rounds, and different 11-round sets produce different ETPs which are averaged up. The ETP is then normalized against the maximum possible ETP achievable by an $N$-household community with the mix giving highest ETP, $\text{ETP}_{\max}$, to obtain the ETPI:
\begin{equation}
    \text{ETPI}_{\text{flex}} = \frac{\text{ETP}_{\text{flex}}}{\text{ETP}_{\max}}
    \label{eq:etpi_flex}
\end{equation}
For example, in a 10-household community in our simulations, the $(1{:}9)$ configuration under the flexible policy has the largest possible ETP of 38.052 when all interactions are assumed to be positive. This value is used as  $\text{ETP}_{\max}$ for normalization across various community mixes. The normalization yields a dimensionless index 
in $[0,1]$ enabling consistent comparison across different community compositions 
and evaluation horizons.

\section{Static Policy}
\label{sec:static}

The static policy represents the special case in which prosumer roles are fixed for 
the entire simulation horizon regardless of the per round energy balance. A 
prosumer always appears on the supplier side of the bipartite graph in every round. 
In rounds where $s_j^r \leq 0$, the prosumer is treated as an inactive supplier, 
submits no offer, and participates in no trades. It is never reclassified as a 
consumer and does not gain access to buy energy from other prosumers. 
Figure~\ref{fig:network} demonstrates the difference between the static and flexible 
approaches with an example round of a (5:5) mix where 2 of the 5 prosumers have no 
surplus. Those 2 deficit prosumers remain isolated on the supplier side in the static 
case, while in the flexible case they move to the consumer side and form edges with 
the surplus-holding prosumers.

Under the static policy, the set of tradeable pairs is the fixed set of 
consumer-prosumer pairs:
\begin{equation}
    \mathcal{E}_{\text{static}} = \{(C_i, P_j) : 1 \leq i \leq N_C,\; 1 \leq j \leq N_P\}
\end{equation}
with $E_{\text{static}}^{\text{total}} = N_C \times N_P$.

\subsection{ETP under the Static Policy}

The ETP and ETPI under the static policy follow the same structure as the flexible 
case but are computed only over the fixed edge set $\mathcal{E}_{\text{static}}$, 
which contains consumer-prosumer pairs exclusively. No prosumer-prosumer pairs exist 
under the static policy, so the second term in \eqref{eq:etp_flex} vanishes:
\begin{equation}
    \text{ETP}_{\text{static}} = \sum_{i=1}^{N_C}\sum_{j=1}^{N_P} T_{ij}^{CP}
    \label{eq:etp_static}
\end{equation}
We get multiple ETP values (one over each trading window of W=11 rounds) and average them all out.

\begin{equation}
    \text{ETPI}_{\text{static}} =
    \frac{\text{ETP}_{\text{static}}}{\text{ETP}_{\max}}
    \label{eq:etpi_static}
\end{equation}

The static policy thus provides a conservative lower bound on trading activity, 
counting only pairs that could trade under the most restrictive role assignment.

\subsection{Per-Round Edge Gain and Optimal Role Switching}

In round $r$, only the $m^r$ prosumers with positive surplus participate as active 
suppliers. The remaining $N_P - m^r$ deficit prosumers are inactive. The number of 
active trading edges under the static policy is:
\begin{equation}
    E_{\text{static}}^r = m^r \cdot N_C
    \label{eq:estatic}
\end{equation}

Comparing \eqref{eq:estatic} with \eqref{eq:eflex}, it is clear that 
$E_{\text{flex}}^r \geq E_{\text{static}}^r$ for all $m^r$, with equality only when 
$m^r = 0$ or $m^r = N_P$. The per-round gain from adopting the flexible policy is:
\begin{equation}
    \Delta_E(m^r) = E_{\text{flex}}^r - E_{\text{static}}^r
    = m^r(N - m^r) - m^r N_C
    \label{eq:gain}
\end{equation}
\begin{equation}
    \Delta_E(m^r) = m^r(N_P - m^r)
    \label{eq:gain2}
\end{equation}

Since $0 \leq m^r \leq N_P$, it follows that $\Delta_E(m^r) \geq 0$ with strict 
inequality whenever $0 < m^r < N_P$. The gain vanishes only in the degenerate cases 
$m^r = 0$ (no suppliers) and $m^r = N_P$ (all prosumers are suppliers), where both 
policies are equivalent. For the $(6{:}4)$ mix, $\Delta_E(m^r)$ peaks at 
$\Delta_E = 4$ when $m^r = 2$, and $\Delta_E = 3$ when $m^r \in \{1, 3\}$.

Since $E_{\text{flex}}^r = m^r(N - m^r)$ is a concave quadratic in $m^r$, the 
optimal number of active suppliers is found by solving:
\begin{equation}
    \max_{m^r}\; m^r(N - m^r), \quad \text{subject to}\quad 0 \leq m^r \leq N_P
    \label{eq:opt_problem}
\end{equation}

Differentiating and setting equal to zero gives the unconstrained stationary point 
$m^s = N/2$. Applying the KKT conditions for the bound constraint $m^r \leq N_P$, 
the constrained optimum is:
\begin{equation}
    m^* = \min\!\left(\frac{N}{2},\; N_P\right)
    \label{eq:mstar}
\end{equation}

The number of prosumers that should ideally switch to the consumer role to maximize 
active edges is $n^*_{\text{switch}} = N_P - m^*$, which simplifies to:
\begin{equation}
    n^*_{\text{switch}} = N_P - \min\!\left(\frac{N}{2},\; N_P\right)
    \label{eq:nswitch}
\end{equation}

\textit{Example:} $N_C = 2$, $N_P = 8$, $N = 10$. Then $m^* = \min(5, 8) = 5$ 
and $n^*_{\text{switch}} = 3$. Three prosumers should switch roles, yielding a 
balanced $(5{:}5)$ configuration with $5 \times 5 = 25$ edges, compared to 
$5 \times 2 = 10$ under the static policy. When $N_P < N/2$, the unconstrained 
optimum is never reached and $n^*_{\text{switch}} = 0$, meaning no switching is 
beneficial and keeping all prosumers as suppliers is already optimal.

To characterize the expected benefit in a stochastic setting, we model each prosumer 
as independently having surplus with probability $q$ in any round, so 
$m^r \sim \text{Binomial}(N_P, q)$ with $\mathbb{E}[m] = N_P q$ and 
$\text{Var}(m) = N_P q(1-q)$. The expected per-round edge gain is then:
\begin{align}
    \mathbb{E}[\Delta_E] &= \mathbb{E}[m(N_P - m)] \notag \\
    &= N_P\,\mathbb{E}[m] - \mathbb{E}[m^2] \notag \\
    &= N_P^2 q - N_P q(1-q) - (N_P q)^2 \notag \\
    &= N_P(N_P - 1)\,q(1-q)
    \label{eq:expected_gain_appendix}
\end{align}

This is maximized at $q = 1/2$, yielding a peak expected gain of:
\begin{equation}
    \mathbb{E}[\Delta_E]_{\max} = \frac{N_P(N_P - 1)}{4}
    \label{eq:peak_gain}
\end{equation}

For $N_P = 9$, this gives $9 \times 8 / 4 = 18$ additional active edges per round. 
The quadratic scaling with prosumer count confirms that flexible role switching is 
most beneficial in prosumer-heavy communities.

\section{Simulations}
\label{sec:simulations}

\subsection{Simulation Methodology}

Both the flexible and static simulations are implemented in Python using a 13-segment pipeline. Each trading round spans one hour and there are 11 trading rounds per day from 08:00 to 19:00. The ETP of each day is computed (W=11 rounds per day). 365 ETP values were calculated over 1 year, and averaged out. The $(0{:}10)$ mix is included only for the flexible case since under the static policy all prosumers remain on the supplier side with 
no buyers present, yielding zero trading activity. The $(10{:}0)$ case is excluded entirely since without any prosumers there is no surplus energy to trade under 
either policy.

Hourly energy consumption data is drawn from the PRECON dataset \cite{nadeem2019} 
for a one-year period (June 2018 to May 2019). The dataset provides residential 
load profiles collected in Pakistan, with average monthly energy demand across the 
ten selected households ranging from approximately 352\,kWh to 535\,kWh. Only 
demand values during market operating hours (08:00 to 19:00) are used. Hourly 
solar generation data is obtained from the NREL PVWatts tool for 3\,kW, 5\,kW, 
7\,kW, and 10\,kW PV system specifications, alternating across prosumer households. 
Independent random perturbations of $\pm$10--20\% are applied to each prosumer's 
base profile to simulate inter-household variability in panel orientation, shading,
and inverter efficiency. All values are constrained to be non-negative. Consumer 
bid prices are drawn uniformly from [PKR\,10, PKR\,14]/kWh and prosumer offer 
prices from [PKR\,8, PKR\,14]/kWh, reflecting the spread between retail 
electricity tariff rates and net metering buyback rates in Pakistan during the 
simulation period. The sentiment classification thresholds are set to 
$\theta_1 = -0.5$ and $\theta_2 = 0.5$, and the floor constant to 
$\epsilon = 0.001$.

\subsection{Simulation Results}

\begin{figure*}[!t]
    \centering
    \includegraphics[width=1\linewidth]{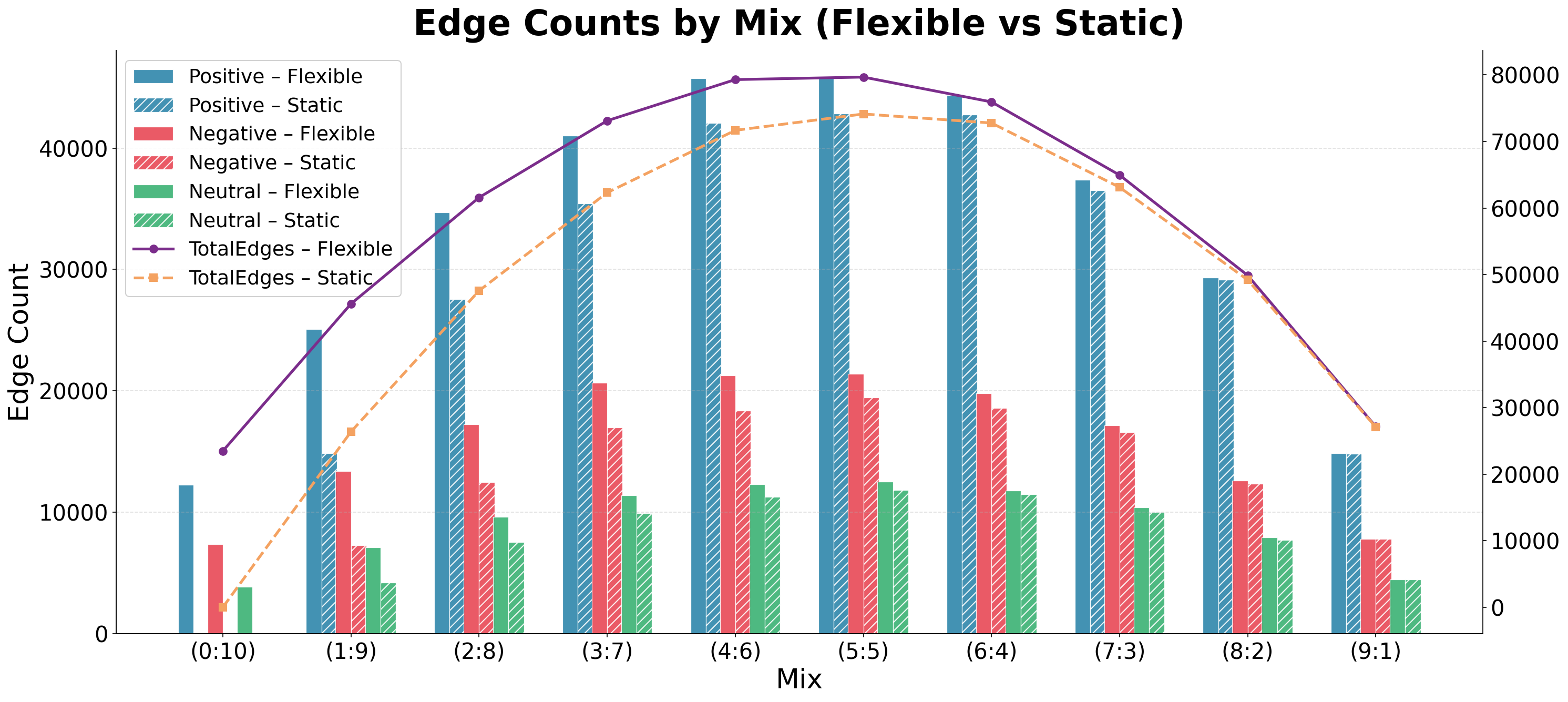}
    \caption{Positive, Negative and Neutral edges count for all the mixes using both the flexible and static policy. The left axis is for the bars, and the right one is for the total edge count lines.}
    \label{fig:edges}
\end{figure*}

\noindent\textbf{Edge Counts:} Figure~\ref{fig:edges} presents the annual active 
edge counts broken down by interaction sentiment alongside the total edge counts 
in both approaches across all ten community mixes. In both policies, positive 
interactions dominate the edge count at every mix. The most pronounced differences 
appear in prosumer-heavy configurations. For the $(1{:}9)$ configuration, the 
flexible policy generates 45,594 total active edges across the simulation year, 
compared to only 26,370 under the static policy, a 73\% increase. With nine 
prosumers, role switching creates up to $9 \times 8 = 72$ possible 
prosumer-prosumer pairs that are entirely absent under the static model. This confirms the analytical result from \eqref{eq:etotal_decomp}. As solar penetration grows and communities become prosumer-heavy, the static model discards 
a substantial portion of realizable trading activity, precisely the scenario 
motivating the flexible policy.

As the mix shifts toward consumer-heavy configurations, the two policies converge. For the $(9{:}1)$ mix, total edges under the two policies are nearly identical (27,144 vs.\ 27,072), consistent with the analytical result from \eqref{eq:gain}. 
With $N_P = 1$ there is only one prosumer to role switch, which does not have a significant impact. The peak total edge count under the flexible policy occurs at the $(5{:}5)$ mix with 79,676 annual active edges, compared to 74,130 for the static policy at the same mix, a more modest 7.5\% difference. The total edge curves form an inverted-U shape as a function of mix in both policies, with the flexible curve lying strictly above the static curve for all prosumer-skewed 
configurations.

\begin{figure}
    \centering
    \includegraphics[width=1\linewidth]{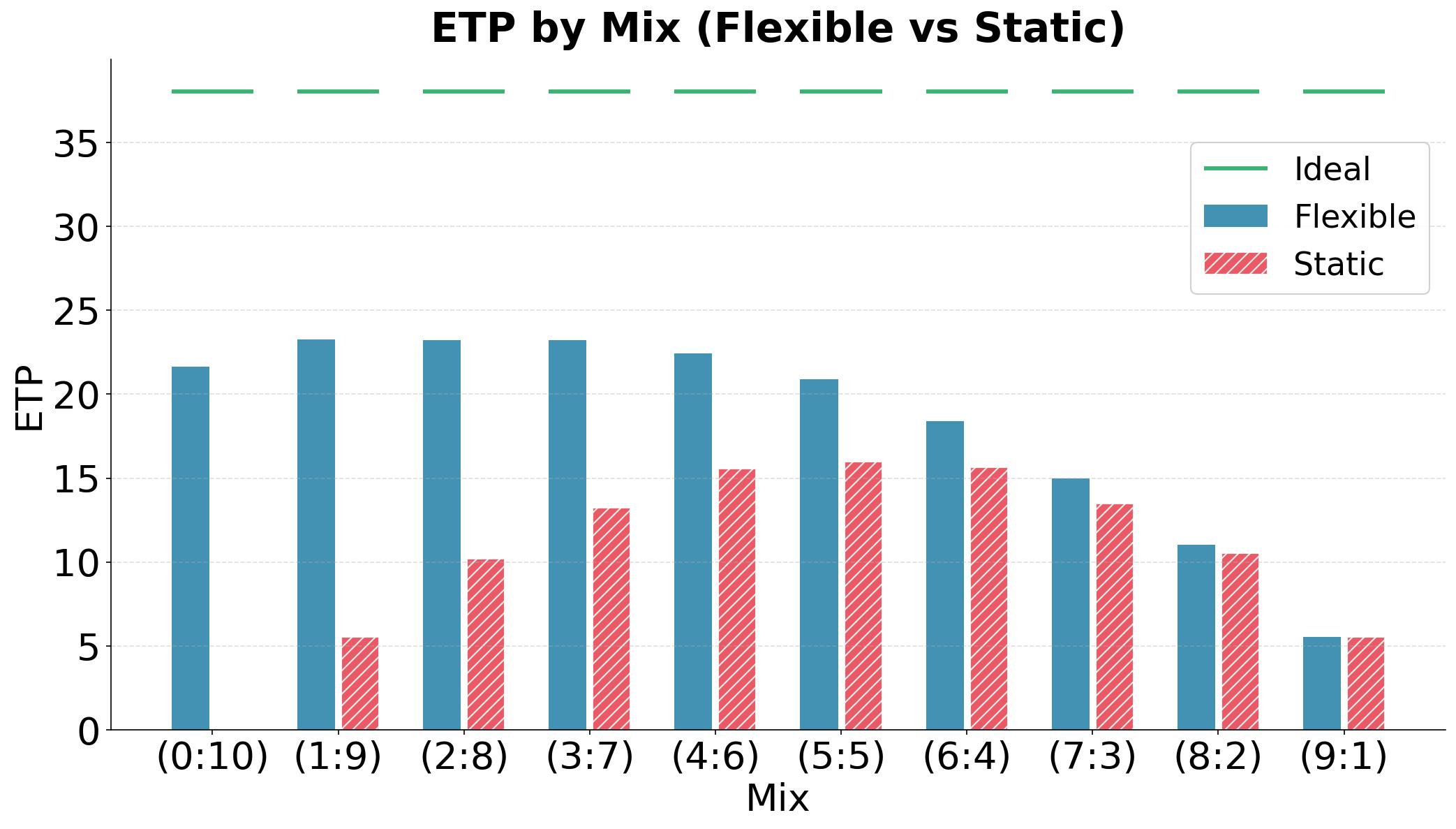}
    \caption{ETP of all the mixes under both the flexible and static policies}
    \label{fig:etp}
\end{figure}

\begin{figure}
    \centering
    \includegraphics[width=1\linewidth]{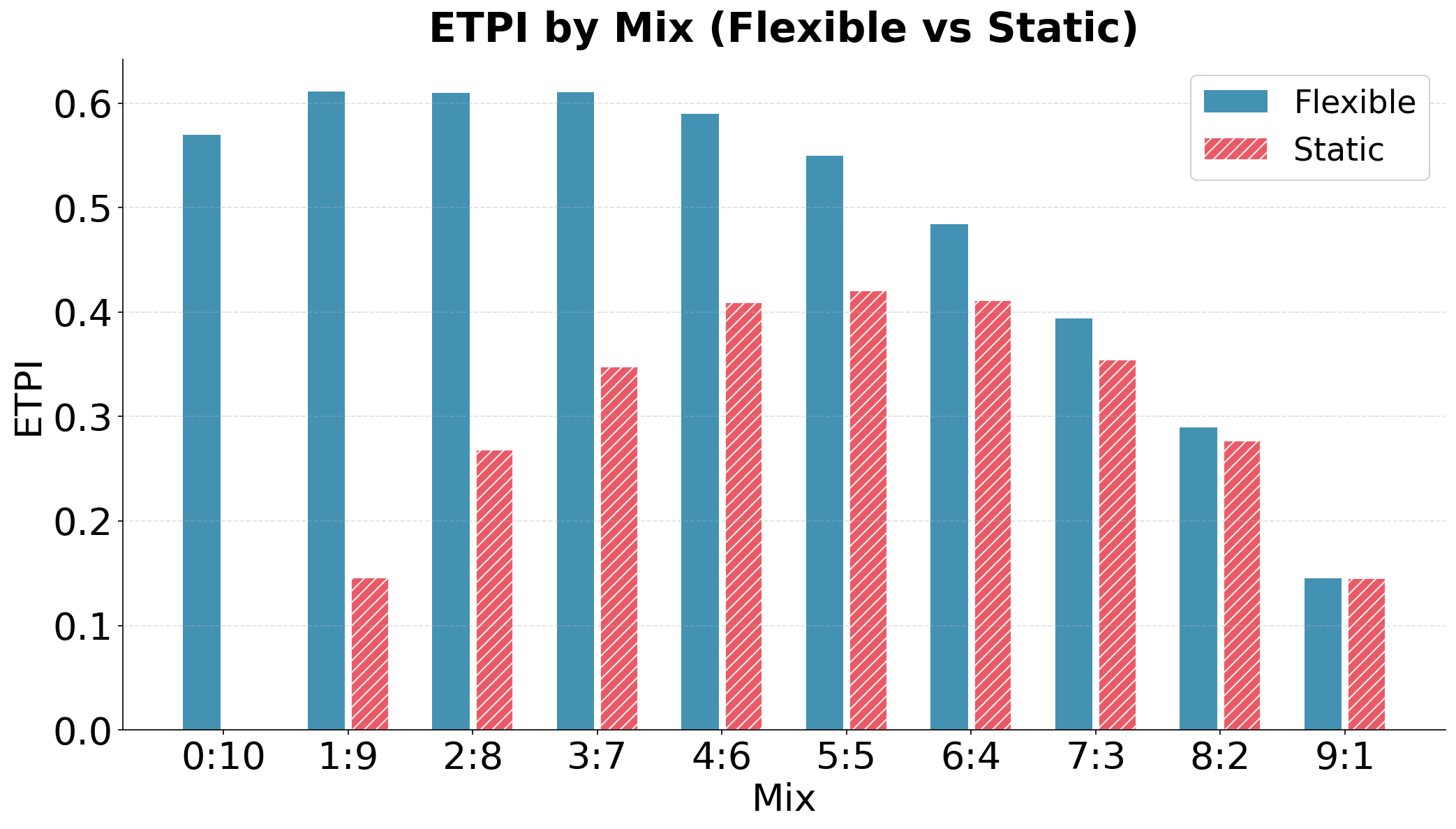}
    \caption{ETPI of all the mixes under both the flexible and static policies}
    \label{fig:etpi}
\end{figure}

\noindent\textbf{ETP:} Figure~\ref{fig:etp} shows the variation of ETP across 
community mixes under both policies. A striking feature of the flexible policy 
is the near-plateau in ETP for prosumer-heavy mixes. The flexible ETP remains 
approximately 22--23 across the $(0{:}10)$ through $(3{:}7)$ configurations 
before declining toward consumer-heavy mixes. This stability arises because role 
switching in prosumer-heavy communities creates a large and diverse pool of active 
pairs, many of which achieve positive or neutral interaction scores, keeping the 
aggregate ETP high regardless of composition. This result directly addresses the open question of whether trading potential 
collapses as neighborhoods become prosumer-heavy. Under the flexible policy it 
does not, even in the extreme $(0{:}10)$ case where all households are prosumers. Notably, although the total edge count peaks at the $(5{:}5)$ mix under both 
policies, the flexible ETP does not. This is because ETP measures the 
quality-weighted sum of trading pairs through $T_{ij}^{CP}$ and $T_{kj}^{PP}$, 
not merely their count. Prosumer-prosumer pairs unlocked by role switching tend 
to achieve higher interaction scores, driving higher edge weights in prosumer-heavy 
configurations even when raw edge count is lower.

The static ETP follows a fundamentally different profile, forming a smooth 
near-symmetric bell curve peaking near the $(5{:}5)$ mix. The static $(1{:}9)$ 
configuration achieves an ETP value of 5.5, which is more than four times lower than the 
flexible $(1{:}9)$ value of $23.3$. This divergence quantifies the trading 
activity discarded by the static model when deficit prosumers are prevented from 
buying surplus energy from peers. A dashed green line at $\text{ETP}_{\max} = 
38.052$ is included for reference, and all realized ETP values lie well below 
this ceiling, highlighting the room for improvement that motivated the ETPI as 
a normalized index.

\noindent\textbf{ETPI:} Figure~\ref{fig:etpi} shows the ETPI across community 
mixes for both policies. Under the flexible policy, the ETPI is highest for 
prosumer-heavy mixes, with values around 0.56--0.62 for the $(0{:}10)$ through 
$(3{:}7)$ range, declining to approximately 0.15 for the $(9{:}1)$ configuration. 
The flexible ETPI does not peak at $(5{:}5)$ but remains elevated across the 
prosumer-heavy range, reflecting the large prosumer-prosumer edge pool enabled 
by role switching. These results have a direct practical implication for communities in markets with declining feed-in tariffs and growing rooftop solar penetration. Such communities 
tend toward prosumer-heavy compositions yet retain substantial P2P trading 
potential and can justify advanced infrastructure investment when assessed under 
the flexible policy.

Under the static policy, the ETPI peaks at the $(5{:}5)$ mix at approximately 
0.42 and falls steeply for both prosumer-heavy and consumer-heavy extremes. The 
static $(1{:}9)$ mix achieves an ETPI of only 0.15, which would incorrectly 
disqualify it from advanced infrastructure investment under a static assessment. 
The flexible ETPI of $\approx 0.61$ for the same mix reveals that the community 
is far more active than the static model suggests, underscoring the importance 
of accounting for role flexibility when evaluating P2P infrastructure deployment.

\begin{figure*}[p]
    \centering
    \includegraphics[width=0.66\textwidth]{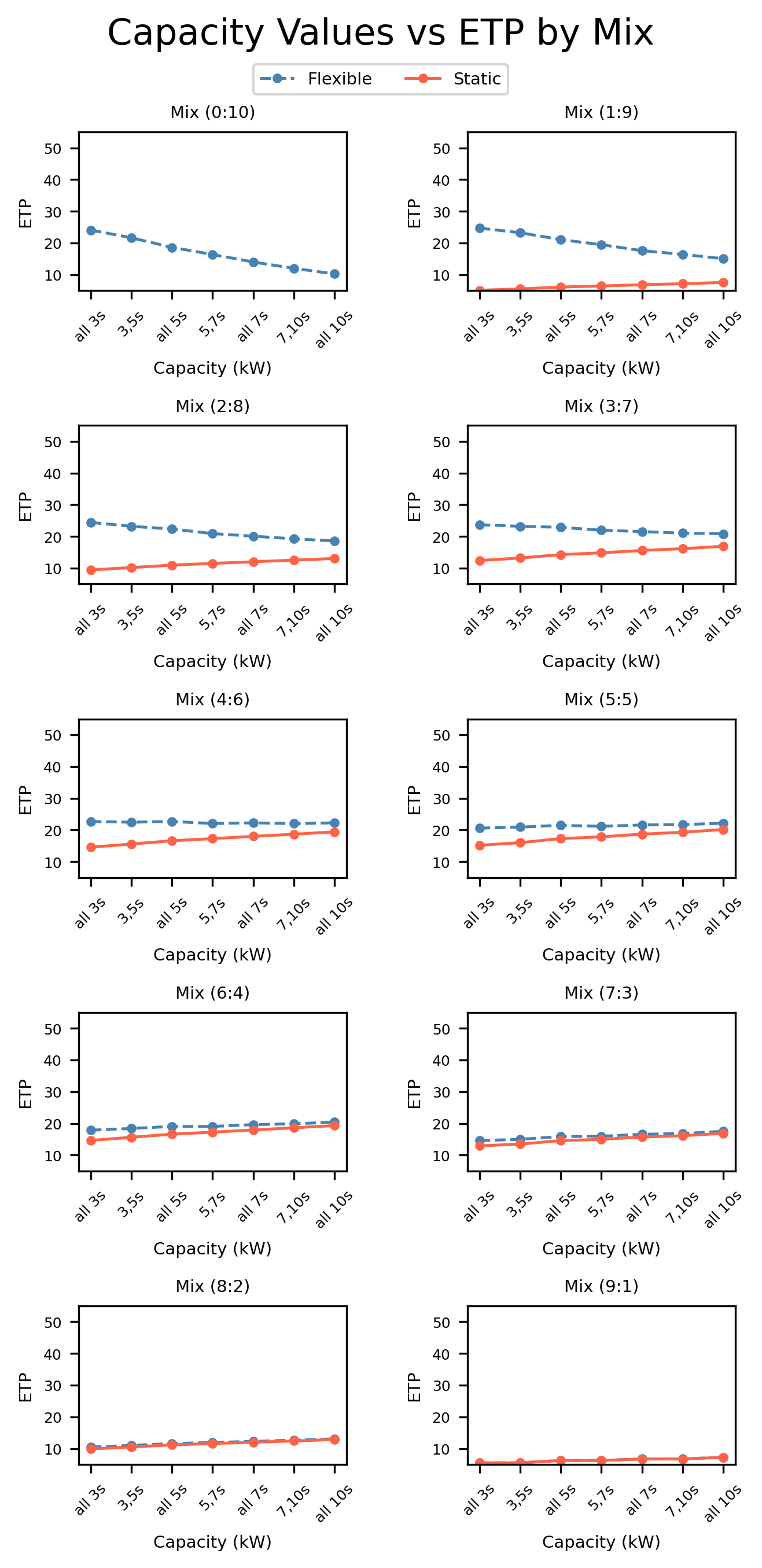}
    \caption{ETP with different capacities for all mixes under both flexible and static policies.}
    \label{fig:CP}
\end{figure*}

\begin{figure*}[p]
    \centering
    \includegraphics[width=0.66\textwidth]{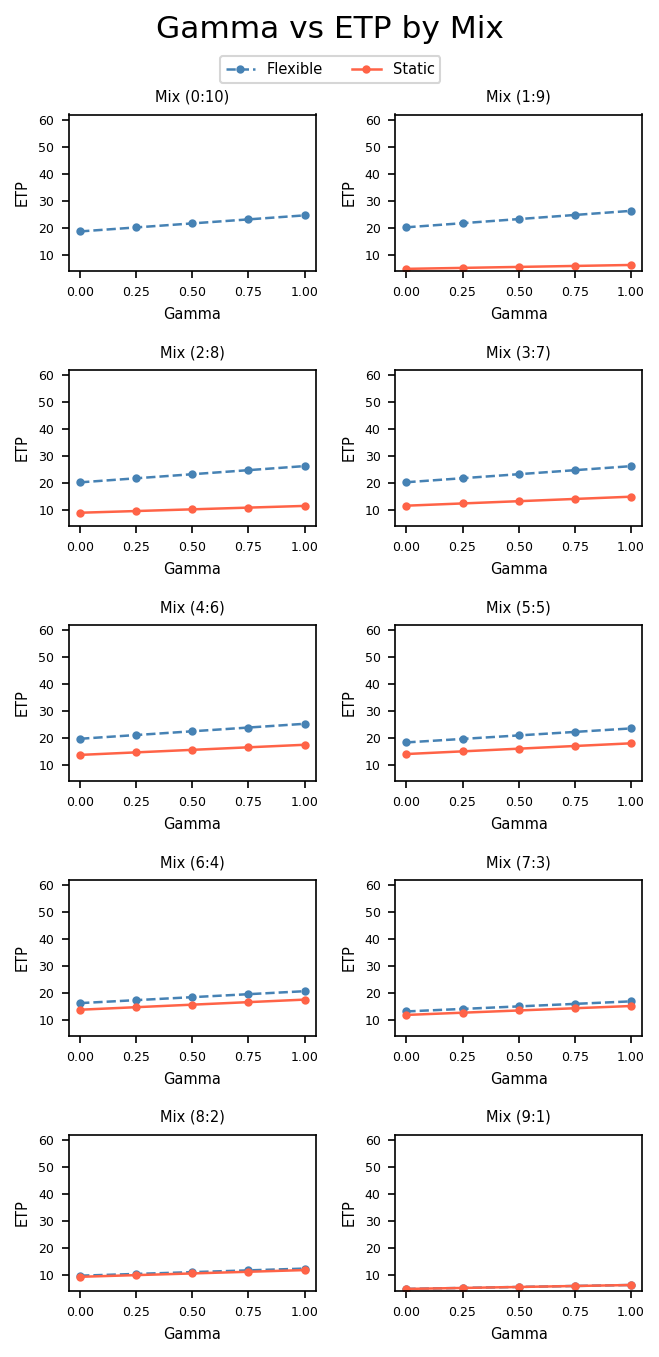}
    \caption{ETP with different gamma values for all mixes using both flexible and static method}
    \label{fig:gamma}
\end{figure*}

\noindent\textbf{ETP vs. PV Capacity:} Figure~\ref{fig:CP} shows how ETP varies with PV system capacity across all mixes under both policies. Six capacity 
configurations are evaluated, ranging from all 3\,kW systems to alternating configurations such as 3\,kW and 5\,kW systems across prosumer households. 
The results reveal a qualitative split between prosumer-heavy and consumer-heavy mixes. For consumer-heavy mixes $(6{:}4)$ through $(9{:}1)$, ETP increases 
monotonically with PV capacity under both policies. Larger solar systems generate more surplus, creating more and larger active trading events. For prosumer-heavy mixes $(0{:}10)$ through $(3{:}7)$, ETP under the flexible policy decreases with 
increasing capacity. This counterintuitive result arises because higher capacity 
systems produce larger surpluses that are rarely matched by the limited demand on 
the consumer side. Larger surpluses also reduce the frequency with which prosumers 
enter deficit and switch to the consumer role, thereby shrinking the prosumer-prosumer edge pool. This finding carries an important practical implication for communities in markets 
with high rooftop solar adoption. Simply maximizing installed PV capacity is not an effective strategy for maximizing P2P trading potential. At balanced mixes such as $(4{:}6)$ and $(5{:}5)$, ETP is relatively 
insensitive to capacity under the flexible policy, reflecting a balance between increased surplus-driven supply and reduced role-switching frequency.

\noindent\textbf{ETP vs. Gamma:} Figure~\ref{fig:gamma} presents the sensitivity of ETP to the hyperparameter $\gamma$ for all mixes under both policies. Increasing $\gamma$ from 0 to 1 monotonically raises ETP for both approaches and across all mixes, as expected from \eqref{eq:edge_weight_cp} and 
\eqref{eq:edge_weight_pp}. Counting neutral interactions as fully positive ($\gamma=1$) inflates the aggregate ETP, while counting them as non-contributory 
($\gamma=0$) deflates it. As the mix shifts toward consumer-heavy configurations, the flexible and static curves converge and nearly overlap at the $(9{:}1)$ mix, 
consistent with the analytical result that the per-round edge gain from role switching vanishes as $N_P$ approaches 1.


\section{Conclusion}
\label{sec:conclusion}

In this paper we proposed an ETPI to quantify the structural impact of flexible role switching on P2P energy trading potential in residential smart grid communities. Simulation results show that prosumer-heavy communities retain substantial trading potential under the flexible role-switching policy. The fourfold ETPI improvement in the $(1{:}9)$ mix directly answers whether trading activity survives as neighborhoods become prosumer-heavy. In markets where declining feed-in tariffs are pushing households toward P2P trading, ETPI provides a principled pre-deployment tool for justifying advanced 
infrastructure investment. Sensitivity analysis further revealed that maximizing installed PV capacity is counterproductive in prosumer-heavy communities, a non-obvious result with direct policy implications. The ETPI framework evolves naturally from a pre-deployment estimate into a live monitoring instrument as trading data accumulates, supporting evidence-based infrastructure decisions throughout the system lifecycle. Future work will explore demand response as an 
active mechanism for improving trading potential by reshaping consumption patterns, and will extend the model to incorporate battery storage, enabling richer 
prosumer behavior and more persistent trading activity.

\bibliographystyle{IEEEtran} 
\bibliography{references}

\end{document}